\documentclass[copyright,creativecommons]{eptcs}
\providecommand{\event}{EXPRESS/SOS 2015} 
\usepackage{breakurl}             

\title{On Compensation Primitives as Adaptable Processes}
\author{Jovana Dedei\'c 
\institute{University of Novi Sad, Serbia}
\and
Jovanka Pantovi\'c
\institute{University of Novi Sad, Serbia}
\and
Jorge A. P\'erez
\institute{University of Groningen, The Netherlands}
}
\def\titlerunning{On Compensation Primitives as Adaptable Processes}
\def\authorrunning{J. Dedei\'c, J. Pantovi\'c \& J.\,A. P\'erez}

\usepackage{color}

\usepackage{color}

\usepackage{amsthm}

\usepackage[inference]{semantic}

\usepackage{color}

\usepackage[utf8]{inputenc}
\usepackage{subfigure}
\usepackage[english]{babel}
\usepackage{amsmath,amsfonts,amssymb}
\usepackage{proof}
\usepackage{graphicx}
\usepackage{multicol}
\usepackage{listings}
\usepackage{url}
\usepackage{cmll}
\usepackage{algpseudocode}
\usepackage{algorithmicx}
\usepackage{enumerate}
\usepackage{textcomp}
\usepackage{latexsym}
\usepackage{epsfig} 
\usepackage{cite} 
\usepackage{xcolor}
\usepackage{bbm}
\usepackage{times}

\usepackage{enumerate}
\usepackage{hyperref}
\usepackage{stmaryrd}
\usepackage{xspace}

\usepackage{packs/mathpartir}
\usepackage{packs/proof-dashed}

\usepackage{makeidx}

\newif{\ifSHORT}
\newif{\iflong}
\newif{\ifLongVersion}
\newif{\ifWithRecords}
\newif{\ifWithProofs}

\newtheorem{theorem}{Theorem}[section]
\newtheorem{lemma}[theorem]{Lemma}
\newtheorem{proposition}[theorem]{Proposition}

\newtheorem{definition}[theorem]{Definition}
\newtheorem{remark}[theorem]{Remark}
\newtheorem{example}[theorem]{Example}

\newcommand{\nil}{{\mathbf{0}}}

\newcommand{\jout}[1]{\overline{#1}}

\newcommand{\para}{\mathord{\;\boldsymbol{|}\;}}
\def\substj#1#2{\{\raisebox{.5ex}{\small$#1$}\! / \mbox{\small$#2$}\}}

\newcommand{\tra}[1]{\xrightarrow{#1}}
\newcommand{\trad}[1]{\xrightarrow{#1}_\mathtt{D}}
\newcommand{\trap}[1]{\xrightarrow{#1}_\mathtt{P}}
\newcommand{\traa}[1]{\xrightarrow{#1}_\mathtt{A}}
\newcommand{\trak}[1]{\xrightarrow{#1}_\kappa}
\newcommand{\redd}{\tra{~~~}}

\newcommand{\namea}{a}
\newcommand{\nameb}{b}

\newcommand{\snamen}{\textit{N}}
\newcommand{\nout}[1]{\overline{#1}}
\newcommand{\contextn}{C}

\newcommand{\contextnd}{D}
\newcommand{\freen}[1]{\mathtt{fn}(#1)}
\newcommand{\boundn}[1]{\mathtt{bn}(#1)}
\newcommand{\freev}[1]{\mathtt{fv}(#1)}
\newcommand{\boundv}[1]{\mathtt{bv}(#1)}

\newcommand{\processr}{R}

\newcommand{\pinactive}{\nil}

\newcommand{\pinst}[2]{\mathtt{inst}\lfloor \lambda #1. #2\rfloor}

\newcommand{\pblock}[1]{\langle #1 \rangle}
\newcommand{\ppblock}[1]{\langle \langle #1 \rangle \rangle}
\newcommand{\pscope}[3]{\ensuremath{#1[#2\,\boldsymbol{,}\,#3]}}
\newcommand{\pscopeb}[3]{\ensuremath{#1\big[#2\,\boldsymbol{,}\,#3\big]}}

\newcommand{\perror}[1]{\overline{#1}}
\newcommand{\finst}[2]{\lambda #1.#2}

\newcommand{\ploc}[2]{#1[#2]}
\newcommand{\plocb}[2]{#1\big[#2\big]}
\newcommand{\plocB}[2]{#1\Big[#2\Big]}
\newcommand{\pupdate}[3]{#1\{(#2).#3\}}

\newcommand{\pupdateB}[3]{#1\Big\{(#2).#3\Big\}}
\newcommand{\kupd}[1]{#1\{\dagger\}}
\newcommand{\idupd}[1]{#1\{{\sf id}\}}

\newcommand{\paupdate}[3]{#1\{#3\}}
\newcommand{\paupdateb}[3]{#1\big\{#3\big\}}

\newcommand{\hole}{\bullet}

\newcommand{\fextraction}[1]{\mathsf{extr}(#1)}
\newcommand{\fextr}[1]{\ensuremath{\fextraction{#1}}}
\newcommand{\fextrd}[1]{\mathsf{extr}_\mathtt{D}(#1)}
\newcommand{\fextrp}[1]{\mathsf{extr}_\mathtt{P}(#1)}
\newcommand{\fextra}[1]{\mathsf{extr}_\mathtt{A}(#1)}
\newcommand{\fextrk}[1]{\mathsf{extr}_\kappa(#1)}

\newcommand{\nocomp}[1]{\mathsf{noComp(#1)}}

\newcommand{\npb}[1]{\mathtt{npb}{(#1)}}

\newcommand{\npbd}[1]{\mathtt{npb}_\mathtt{D}(#1)}

\newcommand{\encd}[2]{\mathtt{{D}}\llbracket #1 \rrbracket_{#2}}
\newcommand{\cencd}[3]{\mathtt{{D}}\Vert #1 \Vert_{#2}^{#3}}
\newcommand{\cencdd}[3]{\mathtt{^d_{D}}\Vert #1 \Vert_{#2}^{#3}}

\newcommand{\npbp}[1]{\mathtt{npb}_\mathtt{P}(#1)}

\newcommand{\encp}[2]{\mathtt{{P}}\llbracket #1 \rrbracket_{#2}}
\newcommand{\cencp}[3]{\mathtt{{P}}\Vert #1 \Vert_{#2}^{#3}}
\newcommand{\cencpd}[3]{\mathtt{^{d}_{P}}\Vert #1 \Vert_{#2}^{#3}}

\newcommand{\npba}[1]{\mathtt{npb}_\mathtt{A}(#1)}

\newcommand{\enca}[2]{\mathtt{{A}}\llbracket #1 \rrbracket_{#2}}
\newcommand{\cenca}[3]{\mathtt{{A}}\Vert #1 \Vert_{#2}^{#3}}
\newcommand{\cencad}[3]{\mathtt{^{d}_{A}}\Vert #1 \Vert_{#2}^{#3}}

\newcommand{\encod}[2]{\llbracket #1 \rrbracket_{#2}}

\newcommand{\nts}[1]{\mathtt{nts}(#1)}

\newcommand{\lts}[1]{\xrightarrow{#1}}
\newcommand{\reduction}{\rightarrow^{*}}

\def\sepr{\; \; \mbox{\Large{$\mid$}}\;\;}

\newdimen\proofrulebreadth \proofrulebreadth=.05em
\newdimen\proofdotseparation \proofdotseparation=1.25ex
\newdimen\proofrulebaseline \proofrulebaseline=2ex
\newcount\proofdotnumber \proofdotnumber=3
\let\then\relax
\def\hfi{\hskip0pt plus.0001fil}
\mathchardef\squigto="3A3B
\newif\ifinsideprooftree\insideprooftreefalse
\newif\ifonleftofproofrule\onleftofproofrulefalse
\newif\ifproofdots\proofdotsfalse
\newif\ifdoubleproof\doubleprooffalse
\let\wereinproofbit\relax
\newdimen\shortenproofleft
\newdimen\shortenproofright
\newdimen\proofbelowshift
\newbox\proofabove
\newbox\proofbelow
\newbox\proofrulename
\def\shiftproofbelow{\let\next\relax\afterassignment\setshiftproofbelow\dimen0 }
\def\shiftproofbelowneg{\def\next{\multiply\dimen0 by-1 }%
\afterassignment\setshiftproofbelow\dimen0 }
\def\setshiftproofbelow{\next\proofbelowshift=\dimen0 }
\def\setproofrulebreadth{\proofrulebreadth}

\def\prooftree{
\ifnum  \lastpenalty=1
\then   \unpenalty
\else   \onleftofproofrulefalse
\fi
\ifonleftofproofrule
\else   \ifinsideprooftree
        \then   \hskip.5em plus1fil
        \fi
\fi
\bgroup
\setbox\proofbelow=\hbox{}\setbox\proofrulename=\hbox{}%
\let\justifies\proofover\let\leadsto\proofoverdots\let\Justifies\proofoverdbl
\let\using\proofusing\let\[\prooftree
\ifinsideprooftree\let\]\endprooftree\fi
\proofdotsfalse\doubleprooffalse
\let\thickness\setproofrulebreadth
\let\shiftright\shiftproofbelow \let\shift\shiftproofbelow
\let\shiftleft\shiftproofbelowneg
\let\ifwasinsideprooftree\ifinsideprooftree
\insideprooftreetrue
\setbox\proofabove=\hbox\bgroup$\displaystyle 
\let\wereinproofbit\prooftree
\shortenproofleft=0pt \shortenproofright=0pt \proofbelowshift=0pt
\onleftofproofruletrue\penalty1
}

\def\eproofbit{
\ifx    \wereinproofbit\prooftree
\then   \ifcase \lastpenalty
        \then   \shortenproofright=0pt  
        \or     \unpenalty\hfil         
        \or     \unpenalty\unskip       
        \else   \shortenproofright=0pt  
        \fi
\fi
\global\dimen0=\shortenproofleft
\global\dimen1=\shortenproofright
\global\dimen2=\proofrulebreadth
\global\dimen3=\proofbelowshift
\global\dimen4=\proofdotseparation
\global\count255=\proofdotnumber
$\egroup  
\shortenproofleft=\dimen0
\shortenproofright=\dimen1
\proofrulebreadth=\dimen2
\proofbelowshift=\dimen3
\proofdotseparation=\dimen4
\proofdotnumber=\count255
}

\def\proofover{
\eproofbit 
\setbox\proofbelow=\hbox\bgroup 
\let\wereinproofbit\proofover
$\displaystyle
}%
\def\proofoverdbl{
\eproofbit 
\doubleprooftrue
\setbox\proofbelow=\hbox\bgroup 
\let\wereinproofbit\proofoverdbl
$\displaystyle
}%
\def\proofoverdots{
\eproofbit 
\proofdotstrue
\setbox\proofbelow=\hbox\bgroup 
\let\wereinproofbit\proofoverdots
$\displaystyle
}%
\def\proofusing{
\eproofbit 
\setbox\proofrulename=\hbox\bgroup 
\let\wereinproofbit\proofusing
\kern0.3em$
}

\def\endprooftree{
\eproofbit 
  \dimen5 =0pt
\dimen0=\wd\proofabove \advance\dimen0-\shortenproofleft
\advance\dimen0-\shortenproofright
\dimen1=.5\dimen0 \advance\dimen1-.5\wd\proofbelow
\dimen4=\dimen1
\advance\dimen1\proofbelowshift \advance\dimen4-\proofbelowshift
\ifdim  \dimen1<0pt
\then   \advance\shortenproofleft\dimen1
        \advance\dimen0-\dimen1
        \dimen1=0pt
        \ifdim  \shortenproofleft<0pt
        \then   \setbox\proofabove=\hbox{%
                        \kern-\shortenproofleft\unhbox\proofabove}%
                \shortenproofleft=0pt
        \fi
\fi
\ifdim  \dimen4<0pt
\then   \advance\shortenproofright\dimen4
        \advance\dimen0-\dimen4
        \dimen4=0pt
\fi
\ifdim  \shortenproofright<\wd\proofrulename
\then   \shortenproofright=\wd\proofrulename
\fi
\dimen2=\shortenproofleft \advance\dimen2 by\dimen1
\dimen3=\shortenproofright\advance\dimen3 by\dimen4
\ifproofdots
\then
        \dimen6=\shortenproofleft \advance\dimen6 .5\dimen0
        \setbox1=\vbox to\proofdotseparation{\vss\hbox{$\cdot$}\vss}%
        \setbox0=\hbox{%
                \advance\dimen6-.5\wd1
                \kern\dimen6
                $\vcenter to\proofdotnumber\proofdotseparation
                        {\leaders\box1\vfill}$%
                \unhbox\proofrulename}%
\else   \dimen6=\fontdimen22\the\textfont2 
        \dimen7=\dimen6
        \advance\dimen6by.5\proofrulebreadth
        \advance\dimen7by-.5\proofrulebreadth
        \setbox0=\hbox{%
                \kern\shortenproofleft
                \ifdoubleproof
                \then   \hbox to\dimen0{%
                        $\mathsurround0pt\mathord=\mkern-6mu%
                        \cleaders\hbox{$\mkern-2mu=\mkern-2mu$}\hfill
                        \mkern-6mu\mathord=$}%
                \else   \vrule height\dimen6 depth-\dimen7 width\dimen0
                \fi
                \unhbox\proofrulename}%
        \ht0=\dimen6 \dp0=-\dimen7
\fi
\let\doll\relax
\ifwasinsideprooftree
\then   \let\VBOX\vbox
\else   \ifmmode\else$\let\doll=$\fi
        \let\VBOX\vcenter
\fi
\VBOX   {\baselineskip\proofrulebaseline \lineskip.2ex
        \expandafter\lineskiplimit\ifproofdots0ex\else-0.6ex\fi
        \hbox   spread\dimen5   {\hfi\unhbox\proofabove\hfi}%
        \hbox{\box0}%
        \hbox   {\kern\dimen2 \box\proofbelow}}\doll%
\global\dimen2=\dimen2
\global\dimen3=\dimen3
\egroup 
\ifonleftofproofrule
\then   \shortenproofleft=\dimen2
\fi
\shortenproofright=\dimen3
\onleftofproofrulefalse
\ifinsideprooftree
\then   \hskip.5em plus 1fil \penalty2
\fi
}

\newcommand{\defref}[1]{Def.\,\ref{#1}\xspace}

\begin{document}
\maketitle

\begin{abstract}
We compare mechanisms for \emph{compensation handling} and \emph{dynamic update} in calculi for concurrency.
These mechanisms are increasingly relevant in the specification of reliable communicating systems.
Compensations and updates are intuitively similar: 
both specify how the behavior of a concurrent system changes at runtime in response to an exceptional event.
However, 
calculi with compensations and updates are technically quite different.
We investigate the \emph{relative expressiveness} of these calculi: 
we develop  encodings of core process languages with compensations
into a calculus of \emph{adaptable processes}  developed in prior work. 
Our encodings  
shed light on the (intricate) semantics of compensation 
handling and its key constructs.  
They also enable the transference of existing verification and reasoning techniques for adaptable processes
to core languages with compensation handling. 
\end{abstract}

\section{Introduction}\label{sec:intro}

Many software applications are based on \emph{long-running transactions} (LRTs). 
Frequently found in service-oriented systems~\cite{DBLP:books/sp/sensoria2011/FerreiraLRVZ11}, LRTs are computing activities which extend in time and may involve distributed, loosely coupled resources. These features sharply distinguish LRTs from usual (database) transactions. One particularly delicate aspect of LRTs management is handling (partial) failures:
mechanisms for detecting failures and bringing the LRT back to a consistent state need to be explicitly programmed. 
As designing and certifying the correctness of such mechanisms is error prone, 
the last decade has
seen the emergence of specialized constructs, such as \emph{exceptions} and \emph{compensations}, which offer direct 
programming support. 
Our focus is in the latter: as their name suggests, compensation mechanisms are meant to compensate the fact that an LRT has failed or has been aborted. Upon reception of an abortion or failure signal, 
compensation mechanisms are expected to install and activate alternative behaviors for recovering system consistency.
Such a compensation behavior may be different from the LRT's initial behavior. 

A variety of calculi for concurrency with constructs for compensation handling has been proposed~(see, e.g.,\cite{BLZ2003,LZ2005,CFV08,DBLP:books/sp/sensoria2011/FerreiraLRVZ11}). Building upon the tradition and approach of mobile process calculi such as the $\pi$-calculus~\cite{DBLP:journals/iandc/MilnerPW92a}, 
they capture different forms of error recovery and offer reasoning techniques (e.g., behavioral equivalences) on communicating processes with compensation constructs.
The relative expressive power of such proposals has also been studied~\cite{CFV08,DBLP:journals/mscs/BravettiZ09,DBLP:conf/esop/LaneseVF10,DBLP:conf/coordination/LaneseZ13}.
On a related but different vein, a calculus of \emph{adaptable processes}
has been put forward as a process calculus approach to 
specify the dynamic evolution of interacting systems~\cite{DBLP:journals/corr/abs-1210-6379}. It is intended as a way 
of overcoming the limitations that process calculi have for describing patterns of dynamic  evolution. 
In this calculus, process behaviors may be enclosed by nested, transparent \emph{locations}; actions of dynamic update  are targeted to particular locations. This model allows us to represent a wide range of evolvability patterns for concurrent processes. 
The theory of adaptable processes includes expressiveness, decidability, and verification results~\cite{DBLP:journals/corr/abs-1210-6379,DBLP:conf/isola/BravettiGPZ12}, as well as the integration with structured communications governed by session types~\cite{GP2015,GP2014}.

Adaptable processes specify forms of dynamic reconfiguration which are triggered by exceptional events, not necessarily catastrophic. For instance, an external request for upgrading a working component is an exceptional event which is hard to predict and entails a modification of the system's behavior. Still, it is certainly not an error or a failure. Thus, 
adaptation intuitively appears to us as a general phenomenon which includes the (negative) exceptional events dealt by compensations. That is, it should be possible to represent failures and compensation activities as particular instances of the behaviors expressible in~\cite{DBLP:journals/corr/abs-1210-6379}. 

In this paper, we make this intuitive observation precise by 
 encoding  calculi with compensations into adaptable processes.
Our motivation is twofold. 
First, given the diversity of linguistic constructs for compensations, understanding how they can be implemented as adaptable processes could shed new light in their formal underpinnings. 
Since adaptable processes have a simple semantics (based on higher-order process communication~\cite{SangiorgiD:expmpa}), 
the envisaged encodings could suggest alternative  semantics for existing formalisms. 
Second, given that adaptable processes have been developed in several directions, 
encodings of calculi with compensations into adaptable processes could enable the transference of, e.g., decidability results or type systems, from
adaptable processes to calculi with compensations.

As source languages in our study, we systematically consider the different classes of calculi with compensations developed in~\cite{DBLP:conf/esop/LaneseVF10},
a work that offers a unified presentation for many calculi proposed in the literature. 
In particular, we consider processes with \emph{static} and \emph{dynamic} compensations, each 
of them with \emph{preserving}, \emph{discarding}, and \emph{aborting} semantics. (All these   semantics are illustrated next.)
As such, we offer six different encodings into adaptable processes, each one equipped with 
appropriate operational correspondence results. The encodings are rather involved; in particular, 
representing preserving, discarding, and aborting
semantics by means of the transparent locations in~\cite{DBLP:journals/corr/abs-1210-6379}
proved to be quite challenging.
In our view, the intricate character of our representations into adaptable processes is directly related to the 
intricate semantics of each of the forms of calculi with compensations.

This paper is structured as follows. 
\S\,\ref{sec:prelim} illustrates primitives for adaptable processes and compensation handling; 
\S\,\ref{sec:calculi} formally presents the corresponding calculi. 
In \S\,\ref{s:staticr} we define and prove  correct encodings of  processes with static compensations into adaptable processes. We consider aborting, preserving, and discarding semantics. 
\S\,\ref{s:dynamicr} describes encodings of processes with dynamic compensations. 
\S\,\ref{s:concl} 
collects some concluding remarks.
Due to space restrictions, 
omitted proofs
can be found online~\cite{DDPonline15}.

\section{Adaptable and Compensable Processes, By Example}\label{sec:prelim}

We give an intuitive account of 
the
calculus of \emph{adaptable processes} (introduced by Bravetti et al.~\cite{DBLP:journals/corr/abs-1210-6379}) and of the 
core calculus with 
primitives for 
\emph{compensation handling} (as presented by Lanese et al.~\cite{DBLP:conf/esop/LaneseVF10,DBLP:conf/coordination/LaneseZ13}).

\paragraph{Adaptable Processes.}
The calculus of \emph{adaptable processes} was introduced in~\cite{DBLP:journals/corr/abs-1210-6379} 
as a variant of Milner's CCS \cite{DBLP:books/daglib/0067019} (without restriction and relabeling), extended with the following two constructs, aimed at representing the dynamic reconfiguration (or update) of active communicating processes:
\begin{enumerate}[1.]
\item A \emph{located process}, denoted $\ploc{l}{P}$, represents a process $P$ which resides in a location called $l$.
Locations are \emph{transparent}: the behavior of $\ploc{l}{P}$ is the same as the behavior of $P$. Locations can also be 
arbitrarily \emph{nested}, which allows to organize process descriptions into meaningful hierarchical structures.
\item An \emph{update prefix} $\pupdate{l}{X}{Q}$---where  $X$ is a process variable 
that occurs zero or more times in $Q$---denotes an adaptation mechanism for processes located at location $l$.
\end{enumerate}
\noindent
This way, in the calculus of adaptable process the possibility of updating a (located) process behavior is given the same status as communication  prefixes. Intuitively, an update prefix for location $l$ is able to interact with a located process at $l$, updating its current behavior. This is captured by the reduction rule
$$
C_1\big[\ploc{l}{P}\big] \para C_2\big[\pupdate{l}{X}{Q}.R \big]
\redd 
C_1\big[Q\substj{P}{X}\big]\para C_2\big[R\big]
$$
where $C_1$ and $C_2$ denote \emph{contexts} which may describe, e.g., nested locations and parallel components.
Therefore, the adaptation mechanism (embodied by $\pupdate{l}{X}{Q}$) moves to the place where $\ploc{l}{P}$ resides ($C_1$ above) and exercises a dynamic update there, as represented by  substitution $Q\substj{P}{X}$. 
As such, adaptation is a form of \emph{higher-order process communication}~\cite{SangiorgiD:expmpa}.
Observe that $Q$ may not contain $X$, so the current behavior at $l$ (i.e., $P$) may get erased as a result of the update.
Notice also that this form of  
adaptation is \emph{subjective}:
located processes are influenced by (unknown) update prefixes in their environment. 

\paragraph{Compensable Processes.}
Our core process language with compensations is based on the calculus in~\cite{DBLP:conf/coordination/LaneseZ13} (a variant of the language in~\cite{DBLP:conf/esop/LaneseVF10}).  
The languages in~\cite{DBLP:conf/coordination/LaneseZ13,DBLP:conf/esop/LaneseVF10} are appealing because they uniformly capture several different proposals for calculi with compensation handling.
These calculi were introduced as extensions of the $\pi$-calculus \cite{DBLP:journals/iandc/MilnerPW92a}  with primitives for \emph{static} and \emph{dynamic recovery}.
However, in order to focus on the essentials of compensation handling primitives, in this presentation we consider a variant of the languages in~\cite{DBLP:conf/coordination/LaneseZ13,DBLP:conf/esop/LaneseVF10} without name mobility. 
There are three salient constructs:
\begin{enumerate}[1.]
\item \emph{Transaction scopes} (or simply \emph{transactions}), denoted $\pscope{t}{P}{Q}$, where $t$ is a name and $P, Q$ are processes;
\item \emph{Protected blocks}, denoted  $\pblock{Q}$, for some process $Q$;
\item \emph{Compensation updates}, denoted $\pinst{X}{Q}.{P}$, where $P, Q$ are processes and $X$ is a process variable that occurs zero or more times in $Q$.
\end{enumerate}
\noindent
While transactions and protected blocks define static recovery mechanisms, 
compensation updates are used to define dynamic recovery. 
We now gradually introduce these constructs and their main features.

\smallskip
\noindent\emph{\underline{Basic Intuitions}.}
A transaction  $\pscope{t}{P}{Q}$ consists of a \emph{default activity} $P$ with a \emph{compensation activity} $Q$. 
Transactions can be nested, so process $P$ in $\pscope{t}{P}{Q}$ may contain other transactions. 
Transactions can be aborted: intuitively, process $\pscope{t}{P}{Q}$ behaves as $P$ 
until an \emph{error notification} (abortion signal) arrives along name~$t$.
Error notifications are simply output messages which can originate inside or outside the transaction. 
To illustrate the simplest manifestation of compensations, we have the following transitions:
$$
\pscope{t}{P}{Q} \para \jout{t}.R \tra{~\tau} Q \para R
\qquad \qquad
\pscope{t}{\jout{t}.P_1 \para P_2}{Q} \para R \tra{~\tau} Q \para R
$$
While the transition in the left shows  how a transaction $t$ can be aborted by an external signal,
the transition in the right illustrate abortion due to an internal signal.
In both cases, abortion leads to discarding the default behavior of the transition, and the compensation 
activity is executed instead ($Q$ in both cases).

\smallskip
\noindent\emph{\underline{Protected Blocks}.}
The  transitions above illustrate the different sources of abortion signals that lead to compensation behaviors.
One key element in  calculi with compensations primitives are \emph{protected blocks}:
as their name suggests, 
these constructs protect a  process from abortion signals. 
Similarly as locations, protected blocks are transparent:
$Q$ and $\pblock{Q}$  have the same behavior, but $\pblock{Q}$ cannot be affected by abortion signals.
Protected blocks are meant 
to prevent abortions after a compensation:
$$
\pscope{t_2}{P_2}{Q_2} \para \jout{t_2} \tra{~\tau} \pblock{Q_2}
$$
That is, the compensation behavior $Q_2$ will be immune to external errors thanks to protected blocks. 
Consider now 
process
$
\pscopeb{t_1}{\pscope{t_2}{P_2}{Q_2} \para \jout{t_2}.R_1 }{Q_1}
$,
which includes a transaction named $t_2$ which is \emph{nested} inside $t_1$.
Although in previous examples the default behavior has been erased following an abortion signal, 
the semantics of compensations actually may partially preserve such behavior. This is realized by 
\emph{extraction functions}, denoted $\fextr{\cdot}$. For the previous process, we have the following transition:
$$
\pscopeb{t_1}{\pscope{t_2}{P_2}{Q_2} \para \jout{t_2}.R_1 \para R_2}{Q_1} \tra{~\tau~} 
\pscopeb{t_1}{\pblock{Q_2} \para \fextr{P_2}\para R_{1}} {Q_1} $$
In case transaction $t_2$ is aborted, its compensation behavior 
 $Q_2$ will be preserved. 
Moreover, part of the behavior of $P_2$ will be preserved as well: this is expressed by process 
$\fextr{P_2}$, which consists of at least
all protected blocks in $P_2$; it 
 may also contain some other processes, related to transactions (see next).
 
 We consider 
 \emph{discarding}, \emph{preserving}, and \emph{aborting} variants for $\fextr{\cdot}$; they define three different  semantics for compensations. 
 Noted $\fextrd{\cdot}$,  $\fextrp{\cdot}$, and $\fextra{\cdot}$, respectively, 
 these functions concern mostly protected blocks and transactions. 
Given a process $P$, we would have:
 \begin{enumerate}[$\bullet$]
\item  $\fextrd{P}$ keeps only protected blocks in $P$. Other processes  (including transactions) are discarded.
\item  $\fextrp{P}$ keeps   protected blocks and transactions at the top-level in $P$. Other processes are discarded.
\item  $\fextra{P}$ keeps protected blocks  and nested transactions in $P$, including their respective compensation 
activities. Other processes are discarded.
\end{enumerate}
As an example, consider the process 
$
P =  \pscopeb{t}{\pscope{t_1}{P_1}{Q_1} \para \pscope{t_2}{\pblock{P_2}}{Q_2} \para R \para \pblock{P_3}}{Q_5}
$.
We then have:
$$
\begin{array}{lrcl}
\text{Discarding semantics:} & \jout{t} \para P & \trad{~\tau~} & \pblock{P_3} \para \pblock{Q_5} \\
\text{Preserving semantics:} &\jout{t} \para P & \trap{~\tau~} & \pblock{P_3} \para \pblock{Q_5} \para \pscope{t_1}{P_1}{Q_1} \para \pscope{t_2}{\pblock{P_2}}{Q_2} \\
\text{Aborting semantics:} & \jout{t} \para P & \traa{~\tau~} & \pblock{P_3} \para \pblock{Q_5} \para \pblock{P_2} \para \pblock{Q_1}  \para \pblock{Q_2}
\end{array}
$$
Thus, the three different semantics implement different levels of protection.
The discarding semantics only concerns the compensation activity for transaction $t$ and the protected block $\pblock{P_3}$.  
The preserving semantics protects also the nested transactions $t_1$ and $t_2$; a process such as $R$, without an enclosing protected block, is discarded.
Finally, the aborting semantics preserves all protected blocks and compensation activities in the default activity for $t$, including those in nested transactions, such as $\pblock{P_2}$.

\smallskip
\noindent\emph{\underline{Dynamic Compensations}.}
Up to here we have considered transactions with \emph{static compensations}: while the default behavior
may change due to transaction abortion, the compensable behavior remains unchanged.
Given a transaction $\pscope{t}{P}{Q}$,
using \emph{compensation updates} one may 
specify in $P$ an 
update for the compensation behavior $Q$.
This is achieved by the operator $\pinst{X}{Q}{.P},$ where $\finst{X}{Q}$ is a function which represents the compensation update. 
As a simple example, consider the following transition:
$$\pscopeb{t}{\pinst{X}{R}. P_1 \para P_2}{Q}  \xrightarrow{~\tau~} \pscopeb{t}{P_1 \para P_2}{R\substj{Q}{X}}  $$ %

\noindent
This way, $\pinst{X}{R}.P$  
produces a new compensation behavior $R\substj{Q}{X}$ after an internal transition. As variable $X$ may not occur in $R$, this step may fully discard the previous compensation activity $Q$.

\section{The Calculi}\label{sec:calculi}

We introduce adaptable processes (\S\,\ref{ss:adap}) and compensable processes (\S\,\ref{ss:comp}).
To focus on 
their essentials, 
both calculi are defined as extensions 
of CCS~\cite{DBLP:books/daglib/0067019} (no name passing involved).
In both cases, we assume a countable set of names $\snamen,$ ranged over by $\namea,\nameb,{l},{t},\ldots.$ 
As a convention, we use names $l, l', \ldots$  
to denote locations (in adaptable processes) and 
names $t, t', \ldots$ to denote
transactions (in compensable processes).

\subsection{Adaptable Processes}\label{ss:adap}

The syntax of the calculus of \emph{adaptable processes} is defined by \emph{prefixes} $\pi, \pi', \ldots$
and \emph{processes} $P, Q, \ldots$:
\begin{eqnarray*}
\pi   ~ ::= ~~ {a} ~\sepr~  \jout{a} ~\sepr~ \pupdate{l}{X}{Q}
\qquad\quad
{P}  & :: = &  \ploc{l}{P}  ~\sepr~ \pinactive ~\sepr~  \pi.{P} ~\sepr~  !{P} ~\sepr~   {P}\para {Q} ~\sepr~  (\nu a){P} ~\sepr~ {X} 
\end{eqnarray*}
We consider  input and output prefixes (noted $a$ and $\jout{a}$, respectively) and the \emph{update prefix} $\pupdate{l}{X}{Q}$, where $Q$ may contain zero or more occurrences of \emph{process variable} $X$.
The syntax of processes  includes
\emph{located processes} (noted $\ploc{l}{P}$ and intuitively motivated above) as well as 
usual CCS constructs for inaction, prefix (sequentiality), replication, parallel composition, and restriction.
We omit $\nil$ whenever possible; we write, e.g., $\pupdate{l}{X}{P}$ instead of $\pupdate{l}{X}{P}.\nil$.
Name $a$ is bound in $(\nu a){P}$
and 
process variable $X$ is bound in $\pupdate{l}{X}{Q}$; given a process $P$,
its sets of free and bound names/variables---denoted $\freen{P}$, $\boundn{P}$, 
$\freev{P}$, and $\boundv{P}$---are  as expected.
We  rely on
expected notions of $\alpha$-conversion (noted $\equiv_{\alpha}$) and process substitution:
$P\substj{Q}{X}$ denotes the 
 process obtained  by (capture avoiding) substitution of $Q$ for $X$ in $P$.


The semantics of adaptable processes is given by a reduction semantics, 
 denoted $ \rightarrow$, and defined as the smallest relation on processes induced by the rules in Figure~\ref{fig:semantictadaptable}.
$\reduction$ denotes the reflexive and transitive closure of $\rightarrow$.
Reduction relies on \emph{structural congruence}, denoted $\equiv$, 
and \emph{contexts}, denoted $C, D, E$.
We define $\equiv$ as 
the smallest congruence on processes that satisfies the axioms:
\[
\begin{array}{lll}
{P} \para {Q} \equiv {Q} \para {P}  & {P} \para ({Q}\para \processr )\equiv ({P} \para {Q})\para \processr & {P} \para \pinactive \equiv {P} \\
{P} \equiv {Q} \text{ if }{P} \equiv_{\alpha} {Q} & (\nu\namea)\pinactive \equiv \pinactive& (\nu\namea)(\nu \nameb){P} \equiv (\nu \nameb)(\nu \namea) {P}  \\ 
(\nu\namea){P} \para {Q} \equiv (\nu \namea)({P}\para {Q}) \text{ if }\namea \notin \freen{Q} & (\nu \namea) \ploc{l}{P} \equiv \ploc{l}{(\nu \namea) P} & ! {P} \equiv {P} \para !{P}  
\end{array} 
\]

\noindent 
The syntax of monadic contexts (processes with a single \emph{hole}, denoted $[\bullet]$) is defined as:
$$
C::= [\bullet]~\sepr~ C  \para P  ~\sepr~ l\big[C \big]
$$

\begin{figure}[t]
\[
\begin{array}{@{\qquad}c@{\qquad}c@{\qquad}c}
\multicolumn{3}{c}{
\inferrule[(R-I/O)]{}{E\Big[C\big[\jout{a}.{P}\big] \para D\big[a.{Q}\big]\Big] 
\rightarrow 
E\Big[C\big[{P}\big]\para D\big[{Q}\big]\Big]} 
\qquad
\inferrule[(R-Upd)]{}{E\Big[C\big[\ploc{l}{P}\big] \para D\big[\pupdate{l}{X}{Q}.R\big]\Big] 
\rightarrow 
E\Big[C\big[Q\substj{P}{X} \big]\para D\big[R\big]\Big] } 
}\vspace{2mm}\\
\inferrule*[left=(R-Par)]{{P} \rightarrow {P'}}{{P} \para Q \rightarrow P' \para {Q}} &
\inferrule*[left=(R-Res)]{{P} \rightarrow {P'}}{(\nu a){P}  \rightarrow (\nu a){P'}} &
\inferrule*[left=(R-Str)]{{P} \equiv {P'}\;\;\; {P'} \rightarrow {Q'} \;\;\; {Q'} \equiv {Q}}{{P}  \rightarrow {Q}} 
\end{array}
\]
\vspace{-5mm}
\caption{Reduction semantics for adaptable processes.\label{fig:semantictadaptable}}
\end{figure}
\noindent We write $C[P]$ to denote the 
process resulting from filling in all occurrences of $[\bullet]$ in context $C$ 
with process $P$.
We comment on rules in Figure~\ref{fig:semantictadaptable}.
Rule \textsc{(R-I/O)} formalizes synchronization between process $ \jout{a}.{P} $ 
and process $a.{Q}$ (enclosed in contexts $C$ and $D$, respectively).
Rule \textsc{(R-Upd)} formalizes the dynamic update/evolvability of a location ${l}$. 
The result of the synchronization between a 
located process $\ploc{l}{P}$ 
and an update prefix $\pupdate{l}{X}{Q}$ 
is the process $Q\substj{P}{X}$. 
This resulting process
stays in the same context as process $\ploc{l}{P}.$ 
Rules \textsc{(R-Par)}, \textsc{(R-Res)}, and \textsc{(R-Str)} are standard and/or self-explanatory.

\subsection{Compensable Processes}\label{ss:comp}

The  calculus of \emph{compensable processes} 
extends CCS with constructs for 
 transactions,
protected blocks, 
and compensation updates:
\begin{eqnarray*}
\pi ~:: =~~ \namea \sepr\nout\namea
\quad\quad~~
{P},{Q} & :: = & \pinactive \sepr  \pi.{P} \sepr !{P} \sepr (\nu a){P} \sepr {P}\para{Q}  \sepr \pscope{t}{P}{Q}  \sepr \pblock{Q} \sepr X \sepr \pinst{X}{R}.P
\end{eqnarray*}
Prefixes $\pi$ include input and output actions.
Processes for inaction ($\pinactive$),  sequentiality ($\pi.P$),
replication ($!P$), 
restriction ($(\nu a){P}$), and
parallel composition ($P\para Q$)
are standard.
We omit $\nil$ whenever possible.
Protected blocks $ \pblock{Q}$, transactions $\pscope{t}{P}{Q}$, and compensation updates $\pinst{X}{R}.P$ have been already motivated.
Error notifications are simply output messages; they can be internal (coming from the default activity) or external (coming from outside of the transaction). Name $a$ is bound in $(\nu a){P}$
and variable $X$ is bound in $\pinst{X}{R}$; given a process $P$,
its sets of free and bound names/variables---denoted $\freen{P}$, $\boundn{P}$, 
$\freev{P}$, and $\boundv{P}$---are   as expected.
$\alpha$-conversion (noted $\equiv_{\alpha}$) and substitution $P\substj{Q}{X}$ 
are also as expected. 
We assume that protected blocks and transactions do not appear behind prefixes; 
this is key  to ensure encoding correctness.
We shall say that the sub-calculus without compensation updates $\pinst{X}{R}.P$ is the calculus with \emph{static compensations}; the full calculus will be referred to as the calculus with \emph{dynamic compensations}.
The following definitions apply uniformly to both.


Following~\cite{DBLP:conf/esop/LaneseVF10,DBLP:conf/coordination/LaneseZ13}, 
the semantics of compensable processes is given in terms of a Labeled Transition System (LTS). Ranged over $\alpha, \alpha'$, the set of labels includes $\namea$, $\nout\namea$, $\tau$, and $\finst{X}{Q}$. 
As in CCS, $\namea$ denotes an input action, $\nout\namea$ denotes an output action, and $\tau$ denotes synchronization (internal action).
Label $\finst{X}{Q}$ is associated to compensation updates. 
Formally, we have three different LTSs, corresponding to processes under discarding, preserving, and aborting semantics. 
Therefore, for each $\kappa \in \{\mathtt{D}, \mathtt{P}, \mathtt{A}\}$, we will have an
extraction  function $\fextrk{\cdot}$ and a transition relation $\trak{~\alpha~}$.
The different extraction functions are defined in Fig.~\ref{fig:extnest};
the rules of the LTSs are given in Fig.~\ref{fig:LTS}. 
As a convention, whenever a notion coincides for the three semantics, we shall avoid decorations 
$\mathtt{D}$, $\mathtt{P}$, and $\mathtt{A}$. This way, e.g., by writing $\fextraction{\pblock{P}}= \pblock{P}$
we mean that the 
extraction
function 
for protected blocks
is the same for all three semantics.

\begin{figure}[t]
\[
\begin{array}{l@{\quad\qquad}l@{\quad\qquad}l}   
  \fextrd{\pscope{t}{P}{Q}}= \pinactive    &  \fextrp{\pscope{t}{P}{Q}}=  \pscope{t}{P}{Q}              & \fextra{\pscope{t}{P}{Q}}= \fextra{P} \para \pblock{Q}\\
  \fextraction{\pblock{P}}= \pblock{P}     & \fextraction{P \para Q} = \fextraction{P}\para  \fextraction{Q}     & \fextraction{(\nu a){P}} = (\nu a)\fextraction{P} \\
  \fextraction{!{P}} = \pinactive & \fextraction{\pinst{X}{Q}{.P}}  = \pinactive & \fextraction{\pi.{P}} = \pinactive \\ 
\end{array}
\]
\vspace{-5mm}
\caption{Extraction functions.
\label{fig:extnest}}
\end{figure}

We comment on the rules in Fig.~\ref{fig:LTS}.
Axioms  \textsc{(L-Out)} and \textsc{(L-In)} execute output and input prefixes, respectively.
Rule \textsc{(L-Rep)} deals with replication, while rule \textsc{(L-Par)} allows one parallel component to progress independently. 
Rule \textsc{(L-Res)} is the standard rule for restriction: it states that a transition of process $P$ determines a transition of process $(\nu a) {P}$, where label $\alpha$ provides that the restriction name $a$ does not occur inside $\alpha$. 
Rule \textsc{(L-Comm)} defines communication on $a$.
Rule \textsc{(L-Scope-Out)} allows the default activity $P$ of a transaction to progress, provided that the performed action is not a compensation update and that there is no pending compensation update to be executed. The latter is ensured by 
condition $\nocomp{P}$, defined in~\cite{DDPonline15}: 
the condition is true if and only if process $P$ does not have compensation update which waits for execution.
This means that a compensation update has priority over other transitions; 
that is, 
if process $P$ in transaction $\pscope{t}{P}{Q}$ has a compensation update at top-level then it will be performed before any change of the current state.
Rule \textsc{(L-Recover-Out)} allows an external process to abort a transaction via an output action $\perror{t}$. The resulting process contains two parts: the first part is obtained from the default activity $P$ of the transaction via the appropriate extraction function; the second part corresponds to compensation ${Q}$ which will be executed inside a protected block. 
Similarly, rule \textsc{(L-Recover-In)} handles abortion when the error notification comes from the default activity $P$ of the transaction.
Rule \textsc{(L-Block)} 
essentially specifies that protected blocks are transparent units. 
Observe that the actual semantics of protected blocks is defined via the extraction functions $\fextr{\cdot}$.
The final two rules are peculiar of processes with dynamic compensations:
while rule \textsc{(L-Inst)} performs a compensation update, 
rule \textsc{(L-Scope-Close)} updates the compensation of a transaction.

\begin{figure}[t]
\[
\begin{array}{@{\qquad\qquad}c@{\qquad\qquad}c@{\qquad\qquad}c@{\qquad\qquad}c} 
\multicolumn{4}{c}{
\inferrule[(L-Out)]{}
        {
          \overline{\namea}.{P} \xrightarrow{\overline{\namea}} {P}
        }
        \quad
	\inferrule[(L-In)]{}
        {
          \namea.{P} \xrightarrow{\namea} {P}
        } 
        \quad
         \inferrule[(L-Rep)]{{P} \xrightarrow{\alpha} {P'}}
        {
          !{P} \xrightarrow{\alpha} {P'} \para !{P}
        } 
        \quad
\inferrule[(L-Par)]{{P} \xrightarrow{\alpha} {P'}}
        {
          {P} \para {Q} \xrightarrow{\alpha} {P'} \para {Q}  
        }
        \quad
  \inferrule[(L-Res)]{{P} \xrightarrow{\alpha} {P'} \ \ \alpha \neq a, \overline{a}}
        {
          (\nu a){P} \xrightarrow{\alpha} (\nu a){P'}
        } 
\quad	
\inferrule[(L-Comm)]{{P} \xrightarrow{\namea} {P'} \ \ \ {Q} \xrightarrow{\nout\namea} {Q'}}
        {
          {P} \ | \ {Q} \xrightarrow{\tau} {P'} \ | \ {Q'}  
        }
        }
\\ \\
\multicolumn{4}{c}{
\inferrule[(L-Scope-Out)]{{P} \xrightarrow{\alpha} {P'} \ \ \ 
\alpha \neq \lambda{X}.{Q} \ \ \ \nocomp{P}}
        {
           \pscope{t}{P}{Q}  \xrightarrow{\alpha} \pscope{t}{P'}{Q}  
        } 
\qquad
\inferrule[(L-Recover-Out)]{\nocomp{P}}
        {
           \pscope{t}{P}{Q}  \xrightarrow{{t}} \fextraction{P} \para \pblock{Q} 
        } 
\qquad
\inferrule[(L-Recover-In)]{{P} \xrightarrow{\nout{t}} {P'} \ \ \ \ \nocomp{P}}
        {
           \pscope{t}{P}{Q}  \xrightarrow{\tau} \fextraction{P'} \para \pblock{Q} 
        } }
\\\\
\multicolumn{4}{c}{
	\inferrule[(L-Block)]{{P} \xrightarrow{\alpha} {P'}}
        {
          \pblock{P}  \xrightarrow{\alpha}  \pblock{P'}
        }  
        \qquad
	\inferrule[(L-Inst)]{}
        {
           \pinst{X}{Q}.{P}  \xrightarrow{\lambda{X}.{Q}} {P}
        } 
\qquad
	\inferrule[(L-Scope-Close)]{{P} \xrightarrow{\lambda{X}.R\,} {P'}}
        {
          \pscope{t}{P}{Q}  \xrightarrow{\tau}  \pscope{t}{P'}{\,R\substj{Q}{X}}
        }    
        }
\end{array}
\]
\vspace{-6mm}
\caption{LTS for compensable processes. Symmetric variants of \textsc{(L-Par)} and \textsc{(L-Comm)} are omitted. \label{fig:LTS}}
\end{figure}

We find it convenient to define structural congruence ($\equiv$) and contexts also for compensable processes. 
We define $\equiv$ as the smallest congruence on processes that 
includes $\equiv_\alpha$ and 
satisfies the axioms: 
\[
\begin{array}{lll}
P \para Q \equiv Q \para  P                   & P \para  (Q \para  R )\equiv (P \para  Q) \para R                                  & P \para \pinactive \equiv P  \\
(\nu a)(\nu b) P \equiv (\nu b)(\nu a) P \qquad  & (\nu a)P \para Q \equiv (\nu a)(P \para Q)  \text{ if }a \notin \freen{Q}  \qquad & (\nu a)\pinactive \equiv \pinactive\\ 
\ppblock{P} \equiv \pblock{P}                & \pblock{(\nu a)P} \equiv (\nu a) \pblock{P}                                      & \pblock \pinactive \equiv \pinactive \\
\multicolumn{2}{l}{\pscope{t}{(\nu a)P}{Q} \equiv (\nu a) \pscope{t}{P}{Q} \text{ if }{t} \neq a, \, a \notin \freen{Q}}   & (\nu a) \jout{a} \equiv \pinactive 
\end{array} 
\]

\noindent 
An $n$-adic context $C[\bullet_1,\ldots,\bullet_n]$ is obtained from a process by replacing $n$ occurrences of $\pinactive,$ that are neither compensations nor in continuation of prefixes, with indexed holes $[\bullet_1], \ldots, [\bullet_n]$.  
This way, for instance, the syntax of monadic contexts is defined as:
$$
C  ::= [\hole]  ~\sepr~  \pblock{ C } ~\sepr~ \pscope{t}{C }{P} ~\sepr~   P \para C  ~\sepr~   C  \para  P ~\sepr~ (\nu a)C .
$$
\noindent
We write $C[P]$ to denote the 
process resulting from filling in all occurrences of $[\bullet]$ in context $C$ 
with process $P$. The following proposition 
is central to our operational correspondence statements.
\begin{proposition}\label{prop:compensable}
Let $P$ be a compensable process. If ${P} \lts{\tau} {P'}$ then one of the following holds:
\begin{itemize}
\item[a)] ${P} \equiv E[C[\nout\namea.{P_1}] \para D[\namea.{P_2}]]$ and 
                ${P'} \equiv E[C[{P_1}] \para D[{P_2}]],$
\item[b)] ${P} \equiv E[C[\pscope{t}{P_1}{Q}] \para  D[\nout{t}.\processr]]$  and 
              ${P'} \equiv  E[C[\fextr{P_1} \para \pblock{Q} ]\para D[\processr]],$
\item[c)] ${P} \equiv C[\pscope{t}{D[\nout{t}.{P_1}]}{Q}]$ and
               ${P'} \equiv  C[\fextr{D[P_1]} \para \pblock{Q}],$
\item[d)] ${P} \equiv E[{t}[C'[\pinst{X}{R}.P],{Q}]]$
and 
${P'} \equiv  E[\pscope{t}{C'[P]}{\,R\substj{Q}{X}}],$
\end{itemize}
for some contexts $\contextn, \contextn', \contextnd,$ $E,$ processes $P_1,P_2,Q,R,$ and names $a,t.$ 
\end{proposition} 

\section{Encoding Static Compensation Processes}\label{s:staticr}

Here we present encodings 
of processes with static compensations into adaptable processes. 
We consider discarding, preserving and aborting semantics.
We adopt the following abbreviations for update prefixes:
\begin{enumerate}[$\bullet$]
\item $\kupd{t}$ for the update prefix $\pupdate{t}{Y}{\pinactive}$ which ``kills'' location $t,$ together with the process located at $t;$
\item $\paupdate{t}{}{P}$  for the update prefix $\pupdate{t}{Y}{P}$ (with $Y \not\in \freev{P}$) that replaces the current behavior at $t$ with $P$; 
\item $\idupd t$ for the update prefix  $\pupdate{t}{X}{X}$ which deletes the location name $t$;
\item $\pupdate{t}{X_1,X_2,\ldots,X_n}{R}$ for the sequential composition of updates $\pupdate{t}{X_1}{\pupdate{t}{X_2}{\cdots .\pupdate{t}{X_n}{R}}}.$
\end{enumerate}

\noindent\textbf{Basic Intuitions.}
We describe some commonalities in the encodings we are about to present.
Unsurprisingly, the main challenge to encodability is in representing transactions
$\pscope{t}{P}{Q}$ 
and protected blocks 
 $\pblock{R}$ 
as adaptable processes. Our strategy consists in representing $P$ and $Q$ independently, using located processes.
Since locations are transparent units of behavior, this suffices for encoding $P$.
However, the encoding of $Q$ cannot freely execute unless an abortion signal 
(an output action)
is received. 
Very approximately, 
our encodings of protected blocks and transactions have the following structure:
\begin{eqnarray}
\encod{\pblock{R}}{t,\rho} & = & \plocb{p_{t,\rho}}{\,\encod{R}{\epsilon}\,} \label{eq:pb} \\
\encod{\pscope{t}{P}{Q}}{\rho} &  = & 
\underbrace{\plocb{t}{\,\encod{P}{t,\rho}\,}}_{\text{(a)}} \para 
\underbrace{l_t.\pi_1.\cdots.\pi_k.\plocb{p_t}{\,\encod{Q}{t,\rho}\,}}_{\text{(b)}} \para 
\underbrace{t.\jout{l_t}.K}_{\text{(c)}} \label{eq:enc}
\end{eqnarray}
In our encodings
we use \emph{paths}, finite sequences of names, denoted
$t_1, t_2, \ldots, t_n$.
The empty path is denoted $\epsilon$.
Ranged over $\rho$, paths 
capture the hierarchical structure of nested transactions.
Using paths, for each protected block, we maintain an association with the name of its enclosing transaction.
As such, the encoding of a protected block associated to transaction $t$ will be enclosed in a location $p_t$ (see \eqref{eq:pb} above).
There could be more than one occurrence of such locations, as the transaction's body may contain several protected blocks.
The encoding of transactions, given in \eqref{eq:enc}, consists of three parallel components:
\begin{enumerate}[$\bullet$]
\item Component (a) is a location which contains the encoding of the default activity of the transaction; 
we retain the name of the transaction in the source process. 
\item  Component (b) represents the compensation activity of the transaction. It is given as a located process
at $p_t$, and is protected by a number of prefixes
$\pi_1,\cdots,\pi_k$
including an input prefix $l_t$.
\item Component (c) handles abortion signals.  After synchronizing with an output on $t$, 
it synchronizes with the input on $l_t$ in  component (b). This releases a process $K$ which   ``collects'' all protected blocks in the encoding of $P$ (which occur inside locations named $p_t$)
but also the encoding of the compensation activity $Q$. This collection process may involve synchronizations with 
$\pi_1, \cdots,\pi_k$ in (b). Once all protected blocks have been collected, location $t$ is destroyed.
\end{enumerate}
\noindent
This (very approximate) strategy is used in all of our encodings, with variations motivated by   
discarding, preserving, and aborting semantics. Knowing the number of protected blocks to be collected is crucial in this scheme.
To this end, appropriate counting functions on the default activity $P$ are defined.

The following remark defines some basic conditions on ``reserved names'' used in our encodings:

\begin{remark}\label{rem:names}
Let $t$ be a name, then we know that there are names $l_{t},k_{t},p_{t}$ and $m_{t}$ which are associated with the name $t.$ Also, if $t_{1}\neq t_{2}$ then 
$l_{t_{1}}\neq l_{t_{2}}, k_{t_{1}}\neq k_{t_{2}}, p_{t_{1}}\neq p_{t_{2}}$ and $m_{t_{1}}\neq m_{t_{2}}.$
\end{remark}

\subsection{Discarding Semantics}

Before presenting the encoding, we introduce some auxiliary functions.
First, 
we introduce a function that  counts the number of protected blocks in a process.

\begin{definition}[Number of protected blocks]\label{def:npbd}
Let $P$ be a compensable process. The number of protected blocks in $P$, denoted  by $\npbd{P}$, is defined as follows:
\[
  \npbd{P}=\left\{
  \begin{array}{ll}
      1 & \text{ if }P=\pblock{P_1}\\
      \npbd{P_1}+ \npbd{P_2} & \text{ if } P=P_1 \para P_2\\
      \npbd{P_1} & \text{ if }P=(\nu a)P_1 \\
      0 & \text{ otherwise.}
  \end{array}
  \right.
\]
\end{definition}

\noindent  
We shall define an encoding   $\encd{\cdot}{\rho}$ of compensable processes into adaptable processes,
where $\rho$ is a path (a sequence of location names).
The encoding of transactions
requires an auxiliary encoding, denoted $\cencd{\cdot}{\rho}{n}$, loosely related to  component (b) in \eqref{eq:enc}. 
In case of an abortion signal $\bar{t}$, $\cencd{\cdot}{\rho}{n}$ defines a process that
collects the encodings of the $n$ protected blocks included in the default activity
(which is to be found at $\rho$)
as well as the encoding of the compensation activity. 
We define $\cencd{\cdot}{\rho}{n}$ by induction on $n$:

\begin{definition}[Auxiliary Encoding]\label{def:auxencodingD}
Let 
$Q$ be a compensable process and let
$\rho_0 = t,\rho$ be a path. Also, let $n \geq 0$. The process
$\cencd{Q}{\rho_0}{n}$ is defined 
as follows:
\[
\begin{array}{ll}
\cencd{Q}{t,\rho}{0}  = & l_t.\jout{m_t}.\plocb{p_\rho}{\encd{Q}{\epsilon}} \para m_t.\jout{k_t}.\kupd{t} \quad \\
\cencd{Q}{t,\rho}{n}  = & l_t.\pupdateB{p_{t,\rho}}{X_1, \cdots, X_n}{\paupdateb{z}{}{\ploc{p_\rho}{X_1}  \para \cdots \para \ploc{p_\rho}{X_n} \para \jout{m_t}.\plocb{p_\rho}{\encd{Q}{\epsilon}}}}.(\ploc{z}{\pinactive} \para m_t.\jout{k_t}.\kupd{t} ) ~~[n>0]
\end{array}
\]
\end{definition}

\noindent 
(The definition of $\encd{\cdot}{\rho}$ is given next.)
Consider the encoding of $\pscope{t}{P}{Q}$: if $P$ contains $n$ top-level protected blocks, 
then
process $\encd{\pscope{t}{P}{Q}}{\rho}$ 
  will include $n$ successive update prefixes
that will look for  $n$ protected blocks at location $p_{t,\rho}$ (the path points that they were enclosed with $t$) and move them to their parent location $p_{\rho}$. As these $n$ dynamic updates leave these located processes at location $t$, an update on $z$  is introduced to take them out of $t$ once the $n$ updates are executed.

We are now ready to introduce the encoding $\encd{\cdot}{\rho}$. Recall that we adhere to Remark~\ref{rem:names}:

\begin{definition}[Encoding Discarding Semantics] \label{def:encodingD}
Let $P$ be a compensable process and let $\rho$ be a path.\label{d:encd}
The encoding $\encd{\cdot}{\rho}$ of compensable processes into adaptable processes is defined as follows:
\[
\begin{array}{lllll}
\multicolumn{5}{c}{
\encd{\pblock{P}}{\rho} =   \plocb{p_\rho}{\encd{P}{\epsilon}} 
\qquad\quad
\encd{\pscope{t}{P}{Q}}{\rho}  =  \plocB{t}{\encd{P}{t,\rho}} \para \cencd{Q}{t,\rho}{\npbd{P}} \para t.\jout{l_t}.k_t.
\pinactive
\qquad\quad
\encd{\pinactive}{\rho}  =  \pinactive
}
\vspace{2mm} \\
\encd{P_1 \para P_2}{\rho}  =   \encd{P_1}{\rho} \para \encd{P_2}{\rho} &
\encd{\pi.P}{\rho}  =  \pi.\encd{P}{\rho} &
\encd{!\,P}{\rho}  =  !\,\encd{P}{\rho} &
\encd{(\nu a)P}{\rho}  =  (\nu a)\encd{P}{\rho} &
\end{array}
\]
\end{definition}
\noindent Key cases are encodings of protected blocks and transactions, as motivated earlier. Each protected block is associated with a location $p$ indexed with the path to the protected block. A transaction is encoded as the composition of three processes. The leftmost component  encodes the default activity $P$  preserving the nested structure. In case of an abortion signal on $t$, the rightmost component will execute the middle component by sending message  $\jout{l_t}$. As already explained, this second component will find all the top-level encodings of protected blocks of $P$, moving them to locations $p_{\rho}$ together with the encoding of compensation activity $Q$. 
We may formalize these observations 
using the following lemma: 

\begin{lemma}\label{lemma:reduction_extraD}
Let $\pscope{t}{P}{Q}$ be a transaction with default activity $P$ and compensation $Q$. Then we have:
\begin{eqnarray*}
 \plocb{t}{\encd{P}{t,\rho}} \para \cencd{Q}{t,\rho}{\npbd{P}}\para \jout{l_t}.k_t  & \reduction & \encd{\fextrd{P}}{\rho} \para \encd{\pblock Q}{\rho}.
\end{eqnarray*}
\end{lemma}


\noindent 
The following statement
attests the operational correspondence for our encoding: 
\begin{theorem}\label{prop:completenessd}
Let $P$ be a compensable process and let $\rho$ be an arbitrary path.
\begin{itemize}
\item[a)] If ${P}\trad{\tau} {P'}$
then
$\encd{P}{\rho} \reduction  \encd{P'}{\rho}.$ 


\item[b)] If  $\encd{P}{\rho} \rightarrow Q$
then there is $P'$ such that $P\trad{\tau} P'$ and $Q\reduction \encd{P'}{\rho}.$

\end{itemize}
\end{theorem}
\smallskip
\noindent We illustrate our encoding by means of an example:

\begin{example}\label{exp:exampleD}
Let $P_0 = \pscope{t}{R \para \pblock{P}}{Q}\para \jout{t}$ be 
a compensable process with $\npbd{R}=0.$
Then $P_0 \trad{~\tau~} \pblock{P}\para\pblock{Q}$. 
By expanding \defref{def:encodingD}, we obtain 
 (recall that we omit $\nil$ whenever possible):\\\\
$\begin{array}{r@{\;}c@{\;}l}
\encd{P_0}{\epsilon} & = & 
 \plocB{t}{\encd{R \para \pblock{P}}{t,\epsilon}} \para \cencd{Q}{t,\epsilon}{1} \para t.\jout{l_{t}}.k_{t}  \para \jout{t} \\
& = & \plocB{t}{\encd{R}{t,\epsilon} \para \plocb{p_{t,\epsilon}}{\encd{P}{\epsilon}}} \para l_t.\pupdateB{p_{t,\epsilon}}{X}{\paupdateb{z}{}{\ploc{p_{\epsilon}}{X} \para\jout{m_{t}}.\plocb{p_{\epsilon}}{\encd{Q}{\epsilon}}} }.(\ploc{z}{\pinactive} \para m_{t}.\jout{k_{t}}.\kupd{t}) \para t.\jout{l_{t}}.k_{t} \para \jout{t} \\
&\!\!\!\!\!\!\reduction & \plocB{t}{\encd{R}{t,\epsilon} \para \paupdateb{z}{}{\ploc{p_{\epsilon}}{\encd{P}{\epsilon}}  \para \jout{m_{t}}.\plocb{p_{\epsilon}}{\encd{Q}{\epsilon}}} }\para \ploc{z}{\pinactive} \para m_{t}.\jout{k_{t}}.\kupd{t}  \para k_{t} 
 \reduction \plocb{p_{\epsilon}}{\encd{P}{\epsilon}}\para \plocb{p_{\epsilon}}{\encd{Q}{\epsilon}}\\
& = & \encd{\pblock{P}\para \pblock{Q}}{\epsilon}
\end{array}$
\end{example}

\subsection{Preserving Semantics}



\noindent 
The encoding of compensable processes with preserving semantics is as the previous encoding. 
In this case, since 
the extraction function keeps both protected blocks and top-level transactions (cf. Fig.~\ref{fig:extnest}),
our auxiliary encoding, denoted $\cencp{\cdot}{\rho}{n,m}$, has two parameters:
$n$ denotes protected blocks and $m$ denotes top-level transactions.
We count protected blocks using \defref{def:npbd}; to
 count  transactions we use the following:

\begin{definition}[Number of transactions]
Let $P$ be a compensable process. The number of  transactions which occur in $P$, denoted $\nts{P}$, is defined as follows:\\
\[ {\nts{P} = \left\{
\begin{array}{ll}	
	\nts{P_1}+1 & \text{ if } {P} = \pscope{t}{P_1}{Q_1}\\
	\nts{P_1} + \nts{P_2} &  \text{ if } {P} = {P_1} \para {P_2}\\ 
	\nts{P_1} &  \text{ if }P = (\nu a)P_1 \\
	0 &  otherwise.\ 
\end{array}
\right.} \]
\end{definition}

The encoding of the transaction body $P$ with location $t$ that is nested in location $\beta_{\rho}.$

\noindent 
Before giving the
encoding $\encp{\cdot}{\rho}$,
we define the auxiliary encoding 
$\cencp{\cdot}{\rho}{n,m}$, where 
$\rho$ is a path, $n$ is the number of protected blocks, and $m$ is the number of transactions
in the default activity.

\begin{definition}[Auxiliary Encoding]\label{def:auxencodingP} 
Let 
$Q$ be a compensable process
and let
$\rho_0 = t,\rho$ be a path. 
Also, let $n, m \geq 0$. The process $\cencp{Q}{\rho_0}{n,m}$ is defined as follows:
\begin{eqnarray*}
\cencp{Q}{t,\rho}{0,0} & = & l_t.\jout{m_t}.a.\plocb{p_\rho}{\encp{Q}{\epsilon}} \para m_t.\jout{k_t}.\kupd{t} \\
\cencp{Q}{t,\rho}{1,0} & = & l_t.\pupdateB{p_{t,\rho}}{X_1}{\paupdateb{z}{}{a.\ploc{p_\rho}{X_1} \para \jout{m_t}.\ploc{p_\rho}{\encp{Q}{\epsilon}}}}.(\ploc{z}{\pinactive} \para m_t.\jout{k_t}.\kupd{t}) \\
\cencp{Q}{t,\rho}{0,1} & = & l_t.\pupdateB{\beta_{t,\rho}}{Y_1}{\paupdateb{z}{}{a.\plocb{\beta_\rho}{Y_1}\para\jout{m_t}.\plocb{p_\rho}{\encp{Q}{\epsilon}}} }.(\ploc{z}{\pinactive} \para m_t.\jout{k_t}.\kupd{t}) \\
\cencp{Q}{t,\rho}{n,m} & = & l_t.\pupdateB{p_{t,\rho}}{X_1,\cdots, X_n}{\pupdateB{\beta_{t,\rho}}{Y_1,\cdots, Y_m}{\paupdateb{z}{}{\ploc{p_\rho}{X_1} \para   \cdots \para \ploc{p_\rho}{X_n} \\
&  & \qquad\para a.(\ploc{\beta_\rho}{Y_1} \para \cdots \para \ploc{\beta_\rho}{Y_m}) \para \jout{m_t}.\plocb{p_\rho}{\encp{Q}{\epsilon}}}}}. (\ploc{z}{\pinactive} \para m_t.\jout{k_t}.\kupd{t})~~ [n,m>0]
\end{eqnarray*}
\end{definition}

\noindent 
We may now define the encoding $\encp{\cdot}{\rho}$:

\begin{definition}[Encoding Preserving] \label{def:encodingP}
Let $P$ be a compensable process and let $\rho$ be a path.
The encoding $\encp{\cdot}{\rho}$ of compensable processes into adaptable processes is defined as 
\begin{eqnarray*}
\encp{\pblock{P}}{\rho}  =   \plocb{p_{\rho}}{\encp{P}{\epsilon}} \qquad \qquad
\encp{\pscope{t}{P}{Q}}{\rho}  =  \plocB{\beta_\rho}{\plocb{t}{\encp{P}{t,\rho}} \para \cencp{Q}{t,\rho}{\npbp{P},\nts{P}} \para t.\jout{l_t}.k_t.\jout j}\para j.\idupd{\beta_{\rho}}.\jout a
\end{eqnarray*}
and as a homomorphism for the other operators.
\end{definition}

The following lemma 
formalizes the execution of the encoding:
\begin{lemma}\label{lemma:reduction_extraP}
Let $\pscope{t}{P}{Q}$ be a transaction with default activity $P$ and compensation $Q$. Then we have: 
\begin{eqnarray*}
& & 
\plocB{\beta_\rho}{\plocb{t}{\encp{P}{t,\rho}} \para \cencp{Q}{t,\rho}{\npbp{P},\nts{P}} \para \jout{l_t}.k_t.\jout j}\para j.\idupd{\beta_{\rho}}.\jout a \reduction  \encp{\fextrp{P}}{\rho} \para \encp{\pblock Q}{\rho}.
\end{eqnarray*}
\end{lemma}

We then have the following statement of operational correspondence:

\begin{theorem}\label{prop:completenessp}
Let $P$ be a compensable process and let $\rho$ an arbitrary path.
\begin{itemize}
\item[a)]  
If ${P}\trap{\tau} {P'}$ then 
$\encp{P}{\rho} \reduction  \encp{P'}{\rho}$.


\item[b)] If $\encp{P}{\rho} \rightarrow Q$ 
then there is $P'$ such that $P\trap{\tau} P'$ and $Q\reduction \encp{P'}{\rho}.$ 

\end{itemize}
\end{theorem}

\begin{example}\label{exampleP}
Let $P_0$ be a compensable process in Example~\ref{exp:exampleD} with $R=\pscope{t_1}{P_1}{Q_1}$ and $\npbp{P_1}=\nts{P_1}=0$. In the preserving semantics we have: 
$P_0 
\trap{~\tau~} 
\pscope{t_1}{P_1}{Q_1}\para \pblock{P}\para\pblock{Q}$.
By expanding \defref{def:encodingP}, we obtain:
\\\\
$
\begin{array}{@{\!}l@{\;}c@{\;}l}
\encp{P_0}{\epsilon}
& = & \plocB{\beta_\epsilon}{\plocb{t}{\encp{\pscope{t_1}{P_1}{Q_1} \para \pblock{P}}{t,\epsilon}} \para \cencp{Q}{t,\epsilon}{1,1} \para t.\jout{l_{t}}.k_{t}.\jout j }\para j.\idupd{\beta_{\epsilon}}.\jout a  \para \bar{t}\\
& = & \plocB{\beta_\epsilon}{\plocb{t}{\plocB{\beta_{t,\epsilon}}{M} \para j.\idupd{\beta_{t,\epsilon}}.\bar{a} 
\para \ploc{p_{t,\epsilon}}{\encp{P}{\epsilon}}}  
 \para l_{t}.\pupdateB{p_{t,\epsilon}}{X_1}{\pupdateB{\beta_{t,\epsilon}}{Y_1}{\paupdateb{z}{}{\plocb{p_\epsilon}{X_1}\para a.\plocb{\beta_\epsilon}{Y_1} \\ && \qquad \qquad \qquad \qquad  \para\jout{m_{t}}.\plocb{p_\epsilon}{\encp{Q}{\epsilon}}}  }}
.(\ploc{z}{\pinactive} \para m_{t}.\jout{k_{t}}.\kupd{t} )  \para t.\jout{l_{t}}.k_{t}.\jout j }
 \para j.\idupd{\beta_{\epsilon}}.\jout a \para \jout{t}\\
&\reduction & \plocB{\beta_\epsilon}{\plocb{t}{\paupdateb{z}{}{\plocb{p_\epsilon}{\encp{P}{\epsilon}}\para a.\plocb{\beta_\epsilon}{M}\para \jout{m_{t}}.\plocb{p_\epsilon}{\encp{Q}{\epsilon}}}  \para j.\idupd{\beta_{t,\epsilon}}.\bar{a}  }
 \para \ploc{z}{\pinactive} \para  m_{t}.\jout{k_{t}}.\kupd{t}  
  \para k_{t}.\jout j } \\ && \para j.\idupd{\beta_{\epsilon}}.\jout a \\
&\reduction & \plocB{\beta_\epsilon}{\plocb{t}{\pinactive\para j.\idupd{\beta_{t,\epsilon}}.\bar{a}  } \para \plocb{p_\epsilon}{\encp{P}{\epsilon}}\para  a.\plocb{\beta_\epsilon}{M}\para \plocb{p_\epsilon}{\encp{Q}{\epsilon}} \para \kupd{t}  \para \jout j } 
 \para j.\idupd{\beta_{\epsilon}}.\jout a  \\
& \reduction & \plocb{\beta_\epsilon}{M} \para  \plocb{p_\epsilon}{\encp{P}{\epsilon}} \para \plocb{p_\epsilon}{\encp{Q}{\epsilon}}
\end{array}
$
\\\\
where $M = \plocb{t_1}{\encp{P_1}{t_1,t,\epsilon}} \para \cencp{Q_1}{t_1,t,\epsilon}{0,0} \para t_1.\jout{l_{t_1}}.k_{t_1}.\jout j$.

\end{example}
\subsection{Aborting Semantics}

%

We now discuss the encoding of compensable processes with abortion semantics.  
While preserving the structure of the two encodings already presented, 
in this case the extraction function (cf. Fig.~\ref{fig:extnest}) add some complications.
We need to modify the function that counts the number of protected blocks in a process;
also, collecting encodings of (nested) protected blocks requires so-called \emph{activation processes} 
which capture the hierarchical structure of nested transactions
(cf. \defref{def:actp}).

\begin{definition}[Number of protected blocks]\label{def:npba}
Let $P$ be a compensable process. The number of protected blocks in $P$, denoted  by $\npba{P}$, is defined as follows:
\[
  \npba{P}=\left\{
  \begin{array}{ll}
      1 & \text{ if }P=\pblock{P_1}\\
      \npba{P_1} + 1 & \text{ if } P= \pscope{t}{P_1}{Q_1}\\
      \npba{P_1} + \npba{P_2} & \text{ if } P= P_1 \para P_2\\
      \npba{P_1} & \text{ if }P=(\nu a)P_1 \\
      0 & \text{ otherwise.}
  \end{array}
  \right.
\]
\end{definition}

We now define the auxiliary encoding, denoted $\cenca{Q}{\rho}{n}$. This process, as explained above,
collects all encoded protected blocks of a process, in a case that an error notification is activated.

\begin{definition}[Auxiliary Encoding]\label{def:auxencodingA}
Let  $Q$ be a compensable process 
and let $\rho_0 = t, \rho$ be a path.
Also, let $n \geq 0$. The process $\cenca{Q}{\rho_0}{n}$ is defined as follows:
\begin{eqnarray*}
\!\!\!\cenca{Q}{t,\rho}{0} \!\!&\!\!\!\! = & \!\!\!\!\!\!\!l_t.\jout{m_t}.\plocb{p_\rho}{\enca{Q}{\epsilon}} \para m_t.\jout{k_t}.\kupd{t}.\Gamma_{t,\rho}  \\
\!\!\!\cenca{Q}{t,\rho}{n} \!\!&\!\!\!\! = & \!\!\!\!\!\!\!l_t.\pupdateB{p_{t,\rho}}{X_1,\cdots,X_n}{\paupdate{z}{}{\ploc{p_\rho}{X_1}  \para \!\cdots \!\para \ploc{p_\rho}{X_n} \para \jout{m_t}.\plocb{p_\rho}{\enca{Q}{\epsilon}}}}.(\ploc{z}{\pinactive} \para m_t.\jout{k_t}.\kupd{t}.\Gamma_{t,\rho}) ~ [n>0]
\end{eqnarray*}
where 
$
\Gamma_{t,\rho}  = \pupdate{\gamma_{t_1}}{Z_1}{\ploc{\gamma_{t_1}}{(\nu l_t)(\nu k_t)(Z_1\para l_t. \jout{k_t} )}}. \cdots. 
\pupdate{\gamma_{t_n}}{Z_n}{\ploc{\gamma_{t_n}}{(\nu l_t)(\nu k_t)(Z_n\para l_t. \jout{k_t})}}.\kupd{\gamma_{t}}
$.
\end{definition}

\noindent
To appropriately collect nested protected blocks, we define a so-called \emph{activation process}
that captures the hierarchical structure of nested transactions.
\begin{definition}[Activation Process]\label{def:actp}
Let $St(P)$ denote the \emph{containment structure} of compensable process $P$, i.e.,
the labeled tree 
(with root $t$)
in which nodes are labeled with transaction names and sub-trees capture transaction nesting. 
The \emph{activation process} for $P$, denoted $\mathcal{T}_t(P)$, is the sequential process 
obtained by a post-order search in $St(P)$ in which
the visit to a node labeled $c_i$ adds prefixes $\overline{l_{c_i}}.k_{c_i}$.
\end{definition}

\noindent
This way, e.g., given 
$
P= 
\pscopeb{a}{\pscope{c}{P_1}{Q_2} \para P_2 }{Q_1} \para \pscopeb{b}{P_3 \para \pscopeb{d}{P_4}{Q_4} \para \pscopeb{e}{P_5}{ Q_5}}{Q_3}
$
we will have the activation process
$\mathcal{T}_t(P)
= 
\jout{l_{c}}.k_{c}.\jout{l_{a}}.k_{a}.\jout{l_{d}}.k_{d}.\jout{l_{e}}.k_{e}.\jout{l_{b}}.k_{b}.\jout{l_{t}}.k_{t}\,
$.
%

 Now we have all necessary definitions for introducing of the encoding 
$\enca{\cdot}{\rho}$
of compensable processes into adaptable processes with respect to aborting semantics.
Notice the use of activation processes in the encoding of transactions:

\begin{definition}[Encoding Aborting]\label{def:encodingA}
Let $P$ be a compensable process and let $\rho$ be a path.
The encoding $\enca{\cdot}{\rho}$ of compensable processes into adaptable processes is defined as 
\begin{eqnarray*}
\enca{\pblock{P}}{\rho} & = &  \plocb{p_\rho}{\enca{P}{\epsilon}}\qquad \qquad 
\enca{\pscope{t}{P}{Q}}{\rho}   =   \plocb{t}{\enca{P}{t,\rho}} \para \cenca{Q}{t,\rho}{\npba{P}} \para \plocb{\gamma_t}{t.\mathcal{T}_t(P)}
\end{eqnarray*}
and as a homomorphism for the other operators.
\end{definition}

\noindent 
With respect to previous encodings, 
the encoding for aborting semantics differs in the 
rightmost process of the encoding. In this case, 
the activation process $\mathcal{T}_t(P)$
searches the subtree of the transaction body to activate the middle components of  all nested transactions inside $t$.

The following correctness statements follow the same ideas as in the two previous encodings.
In the sequel, we write $\approx$ to denote a (weak) behavioral equivalence that abstracts from internal transitions
(due to the synchronizations added by the activation process).
 
\begin{lemma}\label{lemma:reduction_extraA}
Let $\pscope{t}{P}{Q}$ be a transaction with default activity $P$ and compensation $Q$. Then we have: 
$$
\plocb{t}{\enca{P}{t,\rho}} \para \cenca{Q}{t,\rho}{\npba{P}} \para \plocb{\gamma_t}{\mathcal{T}_t(P)}
\reduction  \enca{\fextra{P}}{\rho} \para \enca{\pblock{Q}}{\rho}\para \Gamma_{t,\rho} \para \plocb{\gamma_t}{\pinactive}.
$$
\end{lemma}


\begin{theorem}\label{prop:completenessA}
Let $P$ be a compensable process and 
let 
$\rho$ be an arbitrary path.
\begin{itemize}
\item[a)]
If ${P}\traa{\tau} {P'}$ then 
$\enca{P}{\rho} \reduction  \enca{P'}{\rho}.$


\item[b)] 
If $\enca{P}{\rho} \rightarrow Q$ 
then there is $P'$ such that $P\traa{\tau} P'$ and $Q\reduction  Q'$ and $Q'\approx \enca{P'}{\rho}.$ 


\end{itemize}
\end{theorem}


\section{Encoding Dynamic Compensation Processes}\label{s:dynamicr}

We discuss how to extend the previous encodings to account for compensation updates
$\pinst{Y}{R}.P$. Due to space constraints, we only describe  required extensions to 
previously given definitions/statements. 

\paragraph{Discarding Semantics.}
We first have the following extension to \defref{def:npbd}:
\begin{definition}[Number of protected blocks]\label{def:npbdd}
Let $P$ be a compensable process such that $P = \pinst{Y}{R}.P_1$. The number of protected blocks in $P$, denoted  by $\npb{P},$ is equal to $\npb{P_1}.$ 
\end{definition}

The definition of the auxiliary encoding, given in  \defref{def:auxencodingD}, is extended as follows:

\begin{definition}[Auxiliary encoding]\label{def:auxencodingdD}
Let  
$Q$ be a compensable process
and let 
$\rho_0 = t,\rho$ be a path.
Also, let $n \geq 0$. The process
$\cencdd{Q}{\rho_0}{n}$ is defined inductively on $n$ as follows:
\begin{eqnarray*}
\cencdd{Q}{t,\rho}{0} & = & l_t.\jout{m_t}.\plocb{p_\rho}{\ploc{u}{\jout f.\jout{g} }} \para m_t.\jout{k_t}.\kupd{t} \para \ploc{v}{\pupdate{u}{Z}{(Z\para \ploc{v_1}{\encd{Q}{\epsilon}}\para f.\idupd{v_1}.\idupd{v}.g)}}\\
\cencdd{Q}{t,\rho}{n} & = & l_t.\pupdateB{p_{t,\rho}}{X_1,\cdots,X_n}{ \paupdateb{z}{}{\ploc{p_\rho}{X_1}  \para \cdots \para \ploc{p_\rho}{X_n} \para \jout{m_t}.\plocb{p_\rho}{\ploc{u}{\jout f.\jout g }}}}\\
& &\quad\qquad .(\ploc{z}{\pinactive} \para m_t.\jout{k_t}.\kupd{t})\para  \plocb{v}{\pupdate{u}{Z}{(Z\para \ploc{v_1}{\encd{Q}{\epsilon}}\para f.\idupd{ v_1}.\idupd{ v}.g)}}\quad [n>0]
\end{eqnarray*}
\end{definition}

\noindent Based on the above modifications, the 
encoding of processes with dynamic compensations is obtained by extending \defref{def:encodingD} with the following: 
\begin{eqnarray*}
\encd{Y}{\rho} & = & Y \\
\encd{\pinst{Y}{R}.P}{t,\rho}
& = &\ploc{u}{\pinactive} \para \pupdate{ v_1}{Y}{\jout g.\pupdate v X{X\para  \ploc{v}{\pupdate{u}{Z}{(Z\para \\
 & & \qquad  \ploc{v_1}{\encd{R}{\rho}} \para  f.\idupd{ v_1}.\idupd{ v}.g )}}} \para\encd{P}{t,\rho}}.\jout f.( \ploc{v}{\pinactive} \para \ploc{v_1}{\pinactive})
\end{eqnarray*}

We then have the following property: 

\begin{lemma}\label{lemma:dreduction_extraD}
Let $\pscope{t}{P}{Q}$ be a transaction with default activity $P$ and compensation $Q$. Then we have:
$$ 
\plocb{t}{\encd{P}{\rho}} \para \cencdd{Q}{t,\rho}{\npbd{P}}\para \jout{l_t}.k_t  \reduction  \encd{\fextrd{P}}{\rho} \para \encd{\pblock P}{\rho}
$$
\end{lemma}


\begin{lemma} If $R$ is a compensable process such that all free occurrences   of process variable $X$ in it are replaced with a process $Q$ then the following encoding holds:
$
\encod{R\substj{Q}{X}}{\rho} = \encod{R}{\rho}\substj{\encod{Q}{\rho}}{X}.
$
\end{lemma}

Operational correspondence for the extended encoding follows from the following theorem:
\begin{theorem}\label{prop:dcompletenessd}
Let $P$ be a compensable process and let $\rho$ be an arbitrary path.
\begin{itemize}
\item[a)] If ${P}\trad{\tau} {P'}$ 
then there is an adaptable process $P''$ such that
$\encd{P}{\rho} \reduction P'' $ and $ P''\approx \encd{P'}{\rho}.$

\item[b)] If $\encd{P}{\rho} \rightarrow Q$ 
then there is $P'$ such that $P\trad{\tau} P'$ and
$Q\reduction \encd{P'}{\rho}.$
\end{itemize}
\end{theorem}

\paragraph{Preserving Semantics.}

The function that counts the number of protected blocks in $ \pinst{Y}{R}.P$ is the same as in \defref{def:npbdd}, while a function that counts the number of transactions is defined next.


\begin{definition}[Number of transactions]
Let $P$ be a compensable process such that   $P = \pinst{Y}{R}.P_1$. The number of  transactions which occur in $P$, denoted $\nts{P}$, is equal to $\nts{P_1}.$
\end{definition}

We have the following extension of \defref{def:auxencodingP}:
\begin{definition}[Auxiliary encoding]\label{def:auxencodingdP}
Let 
$Q$ be a compensable process
and let
 $\rho_0 = t,\rho$ be a path.
Also, let $n, m \geq 0$. The process $\cencpd{Q}{\rho_0}{n,m}$ is defined as follows:
\begin{eqnarray*}
\cencpd{Q}{t,\rho}{0,0} & = & l_t.\jout{m_t}.a.\plocb{p_\rho}{\ploc{u}{\jout f.\jout g }} \para m_t.\jout{k_t}.\kupd{t} \para  \plocb{v}{\pupdate{u}{Z}{(Z\para \ploc{v_1}{\encp{Q}{\epsilon}}\para f.\idupd{ v_1}.\idupd{ v}.g)}}\\
\cencpd{Q}{t,\rho}{1,0} & = & l_t.\pupdateB{p_{t,\rho}}{X_1}{\paupdateb{z}{}{a.\ploc{p_\rho}{X_1} \para \jout{m_t}.\ploc{p_\rho}{\ploc{u}{\jout f.\jout g }}}}.(\ploc{z}{\pinactive} \para m_t.\jout{k_t}.\kupd{t}) \\
& & \para  \ploc{v}{\pupdate{u}{Z}{(Z\para \ploc{v_1}{\encp{Q}{\epsilon}} \para f.\idupd{ v_1}.\idupd{ v}.g)}}\\
\cencpd{Q}{t,\rho}{0,1} & = & l_t.\pupdateB{\beta_{t,\rho}}{Y_1}{\paupdateb{z}{}{a.\plocb{\beta_\rho}{Y_1}\para\jout{m_t}.\plocb{p_\rho}{\ploc{u}{\jout f.\jout g }}} }.(\ploc{z}{\pinactive} \para m_t.\jout{k_t}.\kupd{t}) \\
& & \para  \plocb{v}{\pupdate{u}{Z}{(Z\para v_1[\encp{Q}{\epsilon}]\para f.\idupd{ v_1}.\idupd{ v}.g)}}\\
\cencpd{Q}{t,\rho}{n,m} & = & l_t.\pupdateB{p_{t,\rho}}{X_1,\cdots, X_n}{\pupdateB{\beta_{t,\rho}}{Y_1,\cdots, Y_m}{\paupdateb{z}{}{\ploc{p_\rho}{X_1} \para \ploc{p_\rho}{X_2} \para \cdots \para \ploc{p_\rho}{X_n} \\
&  &  \para a.(\ploc{\beta_\rho}{Y_1} \para \cdots \para \ploc{\beta_\rho}{Y_m})\para \jout{m_t}.\plocb{p_\rho}{\ploc{u}{\jout f.\jout g  }}}}}.(\ploc{z}{\pinactive} \para m_t.\jout{k_t}.\kupd{t} ) \para  \plocb{v}{\pupdate{u}{Z}{(Z\para \ploc{v_1}{\encp{Q}{\epsilon}} \\
&  & \para f.\idupd{ v_1}.\idupd{ v}.g)}}\qquad [n,m>0]  
\end{eqnarray*}
\end{definition}

We then have the following extended correctness statements:

\begin{lemma}\label{lemma:dreduction_extraP}
Let $\pscope{t}{P}{Q}$ be a transaction with default activity $P$ and compensation $Q$. Then we have:
$$
\plocB{\beta_\rho}{\plocb{t}{\encp{P}{t,\rho}} \para \cencpd{Q}{t,\rho}{\npbp{P},\nts{P}}\para \jout{l_t}.k_t.\jout j}\para j.\idupd{\beta_{\rho}}.\jout a\reduction  \encp{\fextrp{P}}{\rho} \para \encp{\pblock Q}{\rho}
$$
\end{lemma}



\begin{theorem}\label{prop:dcompletenessp}
Let $P$ be a compensable process and let $\rho$ be an arbitrary path.
\begin{itemize}
\item[a)] If ${P}\trap{\tau} {P'}$ 
then there is an adaptable process $P''$ such that
$\encp{P}{\rho} \reduction P'' $ and $ P''\approx \encp{P'}{\rho}.$


\item[b)] If $\encp{P}{\rho} \rightarrow Q$
 then there is $P'$ such that $P\trap{\tau} P'$ and
$Q\reduction \encp{P'}{\rho}.$
\end{itemize}
\end{theorem}
\paragraph{Aborting Semantics.}
The encoding of  processes with dynamic compensations and aborting semantics
is obtained by extending  \defref{def:encodingA} with the 
encodings of process variables   and compensation updates, which are
the same as in discarding and preserving semantics.
The function that counts  protected blocks in compensation updates $\npb{\pinst{Y}{R}.P}$ is   as in   \defref{def:npbdd}.
We require an extension to \defref{def:auxencodingA}:
\begin{definition}[Auxiliary encoding]\label{def:auxencodingdA}
Let 
$Q$ be a compensable process and let $\rho_0 = t,\rho$ be a path.
Also, let $n \geq 0$. The process $\cencad{Q}{\rho_0}{n}$ is defined as follows:
\begin{eqnarray*}
\cencad{Q}{t,\rho}{0} & = & l_t.\jout{m_t}.\plocb{p_\rho}{\ploc{u}{\jout f.\jout g }} \para m_t.\jout{k_t}.\kupd{t}.\Gamma_{t,\rho} \para 
\ploc{ v}{\pupdate{u}{Z}{(Z\para \ploc{v_1}{\enca{Q}{\epsilon}}\para f.\idupd{ v_1}.\idupd{ v}.g)}}\\
\cencad{Q}{t,\rho}{n} & = & l_t.\pupdateB{p_{t,\rho}}{X_1,\cdots,X_n}{ \paupdate{z}{}{\ploc{p_\rho}{X_1} \para \ploc{p_\rho}{X_2} \para \cdots \para \ploc{p_\rho}{X_n} \para \jout{m_t}.\plocb{p_\rho}{\ploc{u}{\jout f.\jout g }}}}\\
& &\qquad.(\ploc{z}{\pinactive} \para m_t.\jout{k_t}.\kupd{t}.\Gamma_{t,\rho})  \para  \ploc{v}{\pupdate{u}{Z}{(Z\para 
\ploc{v_1}{\enca{Q}{\epsilon}}\para f.\idupd{ v_1}.\idupd{ v}.g)}} \quad [n>0]
\end{eqnarray*}
\end{definition}

We then have the following extended correctness statements: 

\begin{lemma} \label{lemma:dreduction_extraA}
Let $\pscope{t}{P}{Q}$ be a transaction with default activity $P$ and compensation $Q.$ We have:
$$
\plocb{t}{\enca{P}{t,\rho}} \para \cencad{Q}{t,\rho}{\npba{P}} \para \plocb{\gamma_t}{\mathcal{T}_t(P)}
\reduction  \enca{\fextra{P}}{\rho} \para \enca{\pblock{Q}}{\rho}\para \Gamma_{t,\rho} \para \plocb{\gamma_t}{\pinactive}.
$$
\end{lemma}


\begin{theorem}\label{prop:dcompletenessa}
Let $P$ be a compensable process and let $\rho$ be an arbitrary path.
\begin{itemize}
\item[a)] If ${P}\traa{\tau} {P'}$ 
then there is an adaptable process $P''$ such that
$\enca{P}{\rho} \reduction P'' $ and $ P''\approx \enca{P'}{\rho}.$


 \item[b)] If $\enca{P}{\rho} \rightarrow Q$
 then there is $P'$ such that $P\traa{\tau} P'$ and
$Q\reduction \enca{P'}{\rho}.$

\end{itemize}
\end{theorem}


\section{Concluding Remarks}\label{s:concl}
We have compared, from the point of view of relative expressiveness, 
two related and yet fundamentally different process models:
 the calculus of \emph{compensable processes} (introduced in \cite{DBLP:conf/esop/LaneseVF10})
 and the calculus of \emph{adaptable processes} (introduced in \cite{DBLP:journals/corr/abs-1210-6379}).
 We provided encodings of processes with static and dynamic compensations (under
 discarding, preserving, and aborting semantics) into adaptable processes. 
 Our encodings not only are a non trivial application of process mobility as present in adaptable processes; they also shed light on the intricate semantics of
 compensable processes.
As encoding  criteria, we have considered compositionality and operational correspondence (up-to weak equivalences),
as   in~\cite{DBLP:journals/iandc/Gorla10}. 
It would be insightful to establish encoding correctness with respect to all the criteria in~\cite{DBLP:journals/iandc/Gorla10}. 

Our study opens several interesting avenues for future work.
Having addressed the encodability of compensable processes into adaptable processes, we 
plan to consider
the reverse direction, i.e., encodings of adaptable processes into compensable processes.
We conjecture that  an encoding  of adaptable process into a language with static compensations does not exist:  compensation updates 
$\pinst{X}{Q}.P$
seem essential to model an update prefix $\pupdate{l}{X}{Q}.P$---the semantics of both constructs induces process substitutions.
Still, 
even by 
 considering a language with dynamic compensations,
 an encoding of adaptable processes is far from obvious, because the semantics
of compensation updates dynamically modifies the behavior of the compensation activity, the inactive part of a transaction. 
Formalizing these (non) encodability claims is interesting future work.
Another promising direction is to cast our encodability results into
variants of adaptable and compensable processes with \emph{session types}:
a candidate for source language could be the 
 typed calculus with \emph{interactional exceptions} developed in~\cite{DBLP:conf/concur/CarboneHY08}; 
as target language, we plan to consider extensions of 
adaptable processes with session types \cite{GP2015,GP2014}. 

\paragraph{Acknowledgements.}
We are grateful to the anonymous reviewers for their comments and suggestions.
This research was partially supported by the EU COST Action IC1201.
 P\'{e}rez is  also affiliated to NOVA  Laboratory for Computer Science and Informatics,  Universidade Nova de Lisboa, Portugal.

\bibliographystyle{eptcs}
\bibliography{referen}

\end{document}